\documentclass[conference]{IEEEtran}
\IEEEoverridecommandlockouts
\usepackage{cite}
\usepackage{graphicx}

\usepackage{float}
\usepackage{xcolor}
\usepackage{comment}
\usepackage[letterpaper, left=0.75in, right=0.75in, top=0.75in, bottom=1.2in]{geometry}
\usepackage{caption}
\usepackage{tabularx}
\usepackage{subcaption}
\usepackage{acro}[=v2]

\usepackage{makecell}
\usepackage{textcomp}
\usepackage{stfloats}
\usepackage{url}

\usepackage{array}
\usepackage{booktabs}
\usepackage{caption}
\usepackage{longtable}

\usepackage[caption=false,font=small]{subfig}
\usepackage{textcomp}
\usepackage{stfloats}
\usepackage{url}
\usepackage{verbatim}
\usepackage{graphicx}
\usepackage{algorithm}
\usepackage{algorithmic}

\usepackage{cite}
\usepackage{amsmath,amssymb,amsfonts}

\usepackage{csquotes}
\usepackage{xcolor}
\usepackage{graphicx}

\usepackage{color,soul}
\usepackage{lipsum}
\usepackage{multirow}
\usepackage{tabularx,booktabs}

\def\BibTeX{{\rm B\kern-.05em{\sc i\kern-.025em b}\kern-.08em
    T\kern-.1667em\lower.7ex\hbox{E}\kern-.125emX}}
\begin{document}

\title{A Demonstration of Self-Adaptive Jamming Attack Detection in AI/ML Integrated O-RAN}

%\title{Demonstration of Self-Adaptive Jamming Attack Detection in AI/ML Integrated 5G O-RAN Networks}
%Autonomous Detection of Jamming Attacks in Dynamic O-RAN Networks\\
%{\footnotesize \textsuperscript{*}Note: Sub-titles are not captured for https://ieeexplore.ieee.org  and should not be used}

\author{%
Md Habibur Rahman\textsuperscript{1,*},
Md Sharif Hossen\textsuperscript{2,*},
Nathan H. Stephenson\textsuperscript{3},
Vijay K. Shah\textsuperscript{2},
Aloizio Da Silva\textsuperscript{1}\\[0.25em]
\textit{\textsuperscript{1} Commonwealth Cyber Initiative, Virginia Tech, VA, USA, Emails: \{mhrahman, aloiziops\}@vt.edu}\\
\textit{\textsuperscript{2} NextG Wireless Lab, North Carolina State University, NC, USA, Emails: \{mhossen, vijay.shah\}@ncsu.edu}\\
\textit{\textsuperscript{3} George Mason University, VA, USA, Email: nstephe7@gmu.edu}%
\vspace{-0.19in}
\thanks{\textsuperscript{*}Both authors contributed equally.}
}

\maketitle

\begin{abstract}
The open radio access network (O-RAN) enables modular, intelligent, and programmable 5G network architectures through the adoption of software-defined networking, network function virtualization, and implementation of standardized open interfaces. %It also facilitates closed-loop control and (non/near) real-time optimization of radio access network (RAN) through the integration of non-real-time applications (rApps) and near-real-time applications (xApps). 
However, one of the security concerns for O-RAN, which can severely undermine network performance, %and subject it to a prominent threat to the security \& reliability of O-RAN networks 
is \textit{jamming attacks}. %To address it, 
This paper presents SAJD -- a self-adaptive jammer detection framework that autonomously detects jamming attacks in AI/ML framework-integrated O-RAN environments without human intervention. The SAJD framework forms a closed-loop system that includes near-real-time inference of radio signal jamming via our developed ML-based xApp, as well as continuous monitoring and retraining pipelines through rApps. In this demonstration, we will show how SAJD outperforms state-of-the-art jamming detection xApp (offline trained with manual labels) in terms of accuracy and adaptability under various dynamic and previously unseen interference scenarios in the O-RAN-compliant testbed. %Specifically, a labeler rApp is developed that uses live telemetry (i.e., KPIs) to detect model drift, triggers unsupervised data labeling, executes model training/retraining using the integrated \& open-source ClearML framework, and updates deployed models on the fly, without service disruption. Experiments on an O-RAN-compliant testbed show that SAJD outperforms state-of-the-art (offline trained with manual labels) jamming detection xApp in accuracy and adaptability under various dynamic and previously unseen interference scenarios.
%\vspace{-1mm}
\begin{comment}
The Open Radio Access Network (O-RAN) enables modular, intelligent, and programmable 5G radio access network (RAN) architectures through software-defined components like xApps and rApps. This paper introduces SAJD, a self-adaptive jammer detection framework that autonomously detects, classifies, and mitigates jamming attacks in O-RAN environments. SAJD forms a closed-loop system by combining real-time inference via xApps with continuous monitoring and retraining through rApps. It uses live telemetry to detect model drift, triggers unsupervised data labeling and model retraining via ClearML, and updates deployed models on the fly, without service disruption. Experiments on a fully O-RAN-compliant testbed show that SAJD outperforms state-of-the-art (offline trained with manual labels) based static ML models in accuracy and adaptability, handling dynamic and previously unseen interference scenarios with resilience.
\end{comment}
\end{abstract}

\vspace{-0.1in}
\section{Introduction}
As 5G networks become central to mission-critical applications \cite{Ref:Intr-Pirayesh}, from battlefield communications and public safety to industrial automation, they are increasingly vulnerable to targeted wireless attacks. Among the most disruptive of these threats is jamming, where adversaries deliberately inject radio signals into the network’s operating band to degrade communication. %Thanks to the growing availability of software-defined radios (SDRs), launching such attacks has become inexpensive and highly customizable, making it possible for adversaries to craft smart jamming strategies that adapt to network conditions in real time.
These attacks are especially dangerous in dynamic environments such as military deployments, where mobility, terrain, and evolving interference patterns make static defenses ineffective. Traditional interference mitigation approaches rely heavily on predefined models that assume relatively stable operating conditions and do not adapt to unseen jamming scenarios.

This is where the open radio access network (O-RAN) architecture provides a valuable opportunity. By decoupling software from hardware and introducing programmable control layers, including near-real-time radio access network (RAN) intelligent controller (near-RT RIC) and non-real-time RAN intelligent controller (non-RT RIC), O-RAN enables fine-grained monitoring and closed-loop control of RAN behavior \cite{Ref:Intr-Arnaz} \cite{Ref:Intr-Marinova}. Applications such as xApps and rApps can be deployed to intelligently analyze telemetry, detect anomalies, and trigger mitigation actions. These capabilities enable adaptive, AI-driven interference management directly embedded into the network control plane.
%Anand et al. This paper proposes MLMCOS, a machine learning-based xApp for 5G O-RAN to mitigate co-tier interference in HetNets. It classifies users based on service priority and SINR, then offloads high-interference users to nearby femtocells. Implemented within the near-real-time RIC, the solution uses CNN for optimal classification accuracy. MLMCOS outperforms traditional methods in enhancing QoE for video, VoIP, and web services.

Several research efforts have explored machine learning (ML)-based interference detection in O-RAN, including xApps that monitor key performance indicators (KPIs) such as signal-to-interference-plus-noise ratio (SINR), reference signal received power (RSRP), and block error rate (BLER) to detect anomalies. However, most existing studies overlook production-grade adaptive ML systems, which leads to degraded accuracy in dynamic jamming scenarios and a heavy reliance on manual annotation and human-driven retraining \cite{Ref:rw_Anand} \cite{Ref:rw_Nathan}.

This article aims to demonstrate \textit{SAJD — a Self-Adaptive Jammer Detection} framework for 5G O-RAN networks, which is designed to detect and classify %and mitigate 
jamming attacks or jamming interference leveraging an AI/ML-enabled fully closed-loop architecture. It leverages the non-RT RIC for continuous telemetry collection (i.e., network KPIs), data sample annotation for interference detection, an ML-based interference detector, and the near-RT RIC for real-time inference and control. The SAJD framework, being a seamlessly integrated AI/ML framework (i.e., ClearML framework), includes an ML model management pipeline that automatically trains or retrains the interference detector using newly labeled data, pushes updated models to deployed xApps, and initiates mitigation actions, all without interrupting service. As part of the SAJD framework, we have developed and deployed three microservice-based applications — a \textit{Labeler rApp}, a \textit{Training manager rApp}, and an \textit{ML-based real-time interference detection xApp}.

%To address this gap, we propose 
%This article aims to demonstrate our proposed and developed \textit{SAJD — a Self-Adaptive Jammer Detection} framework, integrated with AI/ML framework (e.g., ClearML platform), for 5G O-RAN networks %as illustrated in Fig. \ref{fig:sajd}. SAJD is designed to %detect, classify, and mitigate
% to detect %and classify 
% jamming attacks using a fully closed-loop architecture. 
%It leverages the non-RT RIC for continuous telemetry collection, data labeling, \&  managing training/retraining pipeline of ML based detector, and the near-RT RIC for real-time inference and control.
% The SAJD enables a ML based interference detector management pipeline that includes automatic data labeling without human intervention when environment changes, automatic  training/retraining of the interference detectors using newly labeled data and pushes updated ML based interference detector for the inference utilizing the ClearML platform. As part of the SAJD, we have also developed and deployed three microservice based applications — a \textit{Labeler rApp}, a \textit{Training manager rApp}, and an \textit{ML based real-time interference detection xApp}. 
% \vspace{-0.2cm}

\section{The SAJD Framework}
\begin{figure}[t] %[t]
\centerline{\includegraphics[width=0.4\textwidth]{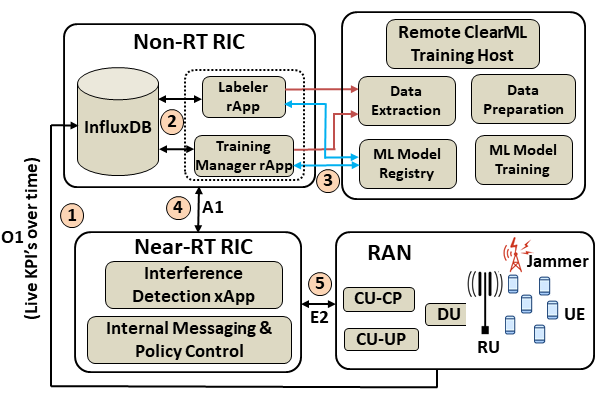}}
\vspace{-0.1in}
\caption{SAJD Closed-Loop Framework.}
\label{fig:sajd}
%\vspace{-5mm}
\end{figure}

Fig.~\ref{fig:sajd} illustrates the overall architecture of the proposed SAJD closed-loop O-RAN system. In this framework, live RAN KPIs such as uplink SNR, MCS, and BLER are continuously collected and stored in the InfluxDB database, then pre-processed and annotated by the \textit{Labeler rApp}. The annotated data is subsequently used by the \textit{Training Manager rApp}, with support from the ClearML platform, to train or retrain a three-layer dense neural network. Finally, the trained ML model is deployed by the \textit{ML-based interference detection xApp} to provide real-time binary decisions (clean or interference), taking new uplink SNR, BLER, and MCS as input for the existing or new scenarios.
 
\section{EXPERIMENTATION AND ANALYSIS}\label{sec:testbed_results}
\begin{figure}[htb]
    \centering
\includegraphics[width=.40\textwidth]{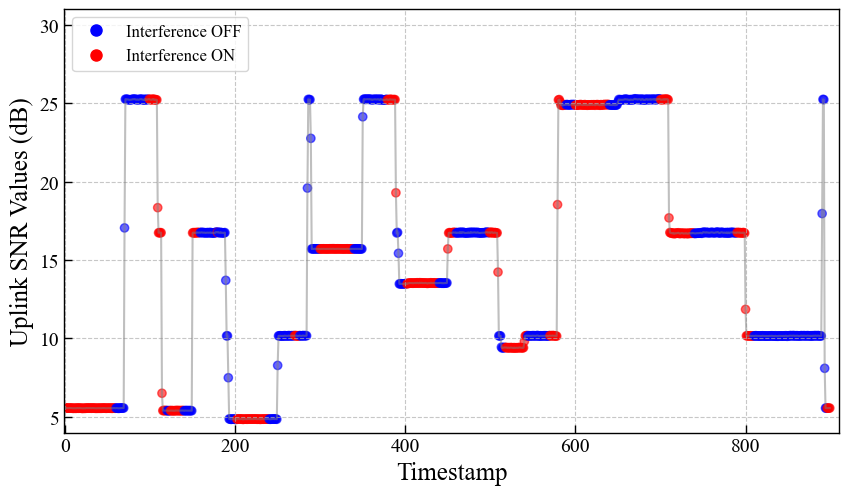}
\vspace{-1mm}
    \caption{Labeler rApp prediction performance for different scenarios.}
    \label{rapp-res}
    \vspace{-3mm}
\end{figure}

\begin{figure}[htbp] %[t]
\centerline{\includegraphics[width=0.35\textwidth]{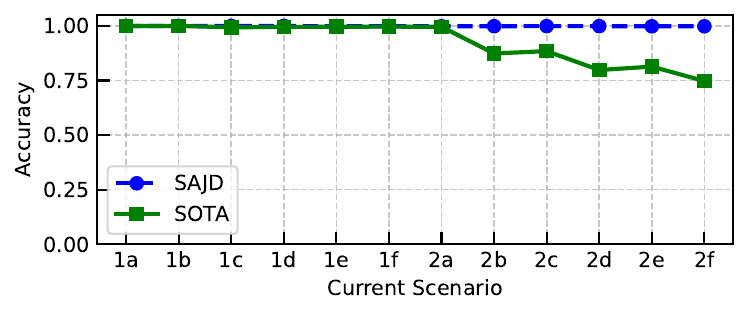}}
\caption{Comparison of interference detection accuracy across windows of sequential scenes (1a–2f). }
\label{fig:accuracy}
\vspace{-4mm}
\end{figure}

%\vspace{-1mm}
%In order 
To evaluate the proposed SAJD framework, we have created an O-RAN-based network using the srsRAN radio protocol stack integrated with the core network as well as O-RAN Software Community near-RT RIC and non-RT RIC components in the NextG Lab \emph{@}NC State \cite{Ref:nglab}. Table \ref{tab:interference_noise} depicts the parameters to train and validate the \textit{Labeler rApp} in the CCI xG Testbed \emph{@}VT \cite{Ref:ccixg}, and Fig. \ref{rapp-res} shows labeling performance, with minor mislabeling near event transitions. Fig. \ref{fig:accuracy} compares the accuracy of the SAJD framework with the state-of-the-art (SOTA) approach across sequential scenario windows in an O-RAN network under dynamically changing interference. While the SOTA approach suffers significant accuracy degradation during transitions into new scenarios, the proposed SAJD approach maintains consistently high accuracy, highlighting SAJD’s superior adaptive effectiveness for interference detection without human intervention.

\begin{table}[t]
\caption{Interference (Int.) and Noise Parameters.}
\label{tab:interference_noise}  
\centering
\begin{tabular}{@{}c c c c c c c c@{}}
\toprule
\textbf{Scen.} & \makecell{\textbf{Int.} \\ \textbf{Event}} & \makecell{\textbf{Int.} \\ \textbf{(dB)}} & \makecell{\textbf{Noise} \\ \textbf{Amp.}} &
\textbf{Scen.} & \makecell{\textbf{Int.} \\ \textbf{Event}} & \makecell{\textbf{Int.} \\ \textbf{(dB)}} & \makecell{\textbf{Noise} \\ \textbf{Amp.}} \\
\midrule
1  & ON   & -8   & 0.056 & 10 & OFF & -100 & 0.15  \\
2  & OFF  & -100 & 0.056 & 11 & ON  & -20  & 0.33  \\
3  & ON   & -8   & 0.15  & 12 & OFF & -100 & 0.33  \\
4  & OFF  & -100 & 0.15  & 13 & ON  & -40  & 0.056 \\
5  & ON   & -8   & 0.33  & 14 & OFF & -100 & 0.056 \\
6  & OFF  & -100 & 0.33  & 15 & ON  & -40  & 0.15  \\
7  & ON   & -20  & 0.056 & 16 & OFF & -100 & 0.15  \\
8  & OFF  & -100 & 0.056 & 17 & ON  & -40  & 0.33  \\
9  & ON   & -20  & 0.15  & 18 & OFF & -100 & 0.33  \\
\bottomrule
\end{tabular}
\vspace{-2mm}
\end{table}

\vspace{-2mm}
\section{Conclusion}\label{sec:conclusion}

In this demonstration, we have introduced SAJD — a self-adaptive jammer detection framework for 5G O-RAN systems that combines ML-driven \textit{Labeler rApp}, \textit{Training manager rApp} \& \textit{ML based
interference detection xApp} and forms a closed-loop pipeline to perform ML-based jamming interference detection in the dynamic network scenarios. Through continuous RAN KPIs collection \& automatic data labeling, ML model training/retraining, and transparent model deployment, the SAJD framework significantly improves interference detection accuracy under previously unseen channel conditions. The findings verify that the SAJD framework is not only intelligently adaptive but also a scalable, efficient, and future-proof interference management solution for future 6G RAN networks. 
\section*{Acknowledgment}
This work has been supported by the NSF awards \#2120411 and \#2443035, and by the Commonwealth Cyber Initiative (CCI)/ CCI xG Testbed. Visit CCI at: \url{www.cyberinitiative.org} and \url{www.ccixgtestbed.org}
%This work was funded by Commonwealth Cyber Initiative (CCI)/ CCI xG Testbed. Visit CCI at: \url{www.cyberinitiative.org} and \url{www.ccixgtestbed.org}.

\bibliographystyle{ieeetr}

\bibliography{main}

\end{document}